\pdfoutput=1
\documentclass[sigconf]{acmart}
\AtBeginDocument{%
  }

\setcopyright{acmlicensed}
\copyrightyear{2018}
\acmYear{2018}
\acmDOI{XXXXXXX.XXXXXXX}
\acmConference[Conference acronym 'XX]{Make sure to enter the correct
  conference title from your rights confirmation email}{June 03--05,
  2018}{Woodstock, NY}
\acmISBN{978-1-4503-XXXX-X/2018/06}
\usepackage{enumitem} %% (1), (2), (3)

\usepackage{xcolor} %%박스
\usepackage{fancybox} %% 박스

\definecolor{rred}{HTML}{FF5834}
\definecolor{ggreen}{HTML}{1AB26B}

\newcommand{\redbox}[1]{%
  \setlength{\fboxsep}{2pt}% 내부 여백
  \fcolorbox{rred}{white}{% 테두리: red, 배경: white
    \textcolor{rred}{\sffamily\small #1}% 글자색: red
  }%
}

\newcommand{\greenbox}[1]{%
  {\color{ggreen}% 테두리 색상을 초록색으로 설정
  \setlength{\fboxsep}{2pt}% 내부 여백
     \colorbox{ggreen}{%
       \textcolor{white}{\sffamily{\small{#1}}}% 흰색 글씨
     }%
   }}%

\begin{document}

\title{AIAP: A No-Code Workflow Builder for Non-Experts with Natural Language and Multi-Agent Collaboration}

\author{Hyunjin An}
\affiliation{%
  \institution{Enhans}
  \city{Seoul}
  \country{Republic of Korea}}
\email{hyunjin@enhans.ai}

\author{Yongwon Kim}
\affiliation{%
  \institution{Department of Animation, \\Kaywon University of Art \& Design}
  \city{Uiwang-si}
  \state{Gyeonggi-do}
  \country{Republic of Korea}}
\email{yongwon@kaywon.ac.kr}

\author{Wonduk Seo}
\affiliation{%
  \institution{Enhans}
  \city{Seoul}
  \country{Republic of Korea}}
\email{wonduk@enhans.ai}

\author{Joonil Park}
\affiliation{%
  \institution{Enhans}
  \city{Seoul}
  \country{Republic of Korea}}
\email{joonil@enhans.ai}

\author{Daye Kang}
\affiliation{%
  \institution{Enhans}
  \city{Seoul}
  \country{Republic of Korea}}
\email{daye@enhans.ai}

\author{Changhoon Oh}
\affiliation{%
  \institution{Graduate School of Information, \\Yonsei University}
  \city{Seoul}
  \country{Republic of Korea}}
\email{changhoonoh@yonsei.ac.kr}

\author{Dokyun Kim}
\affiliation{%
  \institution{Enhans}
  \city{Seoul}
  \country{Republic of Korea}}
\email{dokyun@enhans.ai}

\author{Seunghyun Lee}
\authornote{Corresponding author}
\affiliation{%
  \institution{Enhans}
  \city{Seoul}
  \country{Republic of Korea}}
\email{seunghyun@enhans.ai}


\newcommand{\sysname}[0]{AIAP}

\begin{abstract}
While many tools are available for designing AI, non-experts still face challenges in clearly expressing their intent and managing system complexity. We introduce AIAP, a no-code platform that integrates natural language input with visual workflows. AIAP leverages a coordinated multi-agent system to decompose ambiguous user instructions into modular, actionable steps, hidden from users behind a unified interface. A user study involving 32 participants showed that AIAP's AI-generated suggestions, modular workflows, and automatic identification of data, actions, and context significantly improved participants' ability to develop services intuitively. These findings highlight that natural language-based visual programming significantly reduces barriers and enhances user experience in AI service design.
\end{abstract}



\keywords{Do, Not, Us, This, Code, Put, the, Correct, Terms, for,
  Your, Paper}
\maketitle
\section{Introduction}

% The rapid advancement of artificial intelligence (AI) has democratized access to complex technical capabilities, enabling non-experts to perform sophisticated tasks that previously required specialized knowledge ~\cite{kanbach2024genai}. This democratization extends to AI service development itself, where individuals without extensive technical backgrounds can now create and deploy AI-powered applications. Large Language Models (LLMs) have been particularly transformative, introducing natural language-based interactions that allow users to communicate their intentions through familiar conversational interfaces rather than formal programming languages. Furthermore, multi-agent systems—where specialized AI agents collaborate to solve problems—have significantly enhanced both the scope and accuracy of tasks that non-experts can accomplish.

% Despite these advancements, current chat-based interfaces serving as the primary user touchpoint still present significant limitations. Users often struggle to clearly define objectives and accurately assess system capabilities, frequently resorting to iterative trial-and-error approaches \cite{johnny, subramonyam2024bridging}. This pattern increases cognitive workload and reduces efficiency during the AI service development process, creating a barrier between non-expert users and the full potential of AI technologies.

Recent advancements in artificial intelligence (AI)—particularly in large language models (LLMs) and multi-agent systems—have significantly expanded the range of tasks that can be automated or supported through intelligent systems~\cite{kanbach2024genai}. They enable powerful capabilities such as reasoning over unstructured inputs, coordinating specialized agents, and interpreting open-ended goals. Despite lowering entry barriers, tailoring these technologies to specific goals or integrating them into end-to-end applications still requires substantial technical expertise.

In practice, non-experts encounter significant challenges when building AI-powered applications. Although chat-based interfaces provide a low barrier to entry, their linear and opaque nature limits support for complex service development~\cite{johnny, subramonyam2024bridging}. Users often struggle to articulate high-level goals, break them down into actionable steps, or interpret system feedback, leading to inefficient trial-and-error workflows that increase cognitive load and hinder development efficiency~\cite{johnny}.

Visual programming tools have long aimed to support non-programmers by providing graphical interfaces that enable software development without writing code~\cite{myers1990taxonomies}. However, their original focus on well-defined programming tasks limits their ability to support the ambiguous goals, iterative workflows, and agent-based reasoning involved in AI service creation~\cite{amershi2019guidelines}. Consequently, they offer limited support for interpreting user intent or adapting to evolving requirements, which constrains their suitability for developing modern, conversational AI services.

To address these challenges, we introduce \sysname{} (AI Agent Platform), a no-code development environment designed to empower non-experts to build and deploy sophisticated AI services. \sysname{} combines natural language interaction, visual workflow construction, and multi-agent collaboration in a unified system. Users can express their intentions using everyday language, and refine them through an intuitive modular interface. Internally, specialized AI agents interpret user input, identify relevant components, and connect to appropriate APIs and tools—enabling users to build complete AI services without writing any code.

\sysname{} reduces cognitive load and enhances usability through three core features: (1) \textit{AI-generated suggestions} that convert natural language inputs into structured, actionable steps; (2) a modular, node-based visual interface that facilitates intuitive service construction and debugging; and (3) automatic identification and linking of data, actions, and context via intelligent multi-agent collaboration. This integrated approach retains the expressive power required for sophisticated applications while significantly lowering the barrier to entry.

To evaluate the effectiveness of \sysname{}, we conducted a two-stage user study with non-expert participants. In the first stage, 22 participants completed structured tasks using \sysname{}, and provided initial feedback on usability and system clarity. Based on these insights, we refined the system. In the second stage, 10 participants independently designed and built AI services in a free-form exploration study. The results demonstrated that participants were able to create functional AI applications with minimal guidance, reporting high satisfaction and improved efficiency throughout the development process.

This paper makes the following key contributions to the HCI community:
\begin{itemize}
\item We present the design and implementation of \sysname{}, a no-code AI service development platform that integrates natural language prompts, visual programming, and multi-agent collaboration to support non-expert users.
\item We propose a novel automated pipeline that leverages multi-agent collaboration to extract user intentions (data, actions, context) from natural language and link them to relevant APIs and tools.
\item We empirically validate the usability and effectiveness of our visual programming framework in lowering the barrier to AI service development for non-experts.
\end{itemize}

\section{Related work}
This section reviews the concept of visual programming, explores the integration of LLMs into visual environments, and examines the cognitive challenges associated with natural language-based LLM interfaces. Additionally, we discuss recent advances in LLM-based multi-agent systems and their application in workflow design.

\subsection{Visual Programming}
Visual programming uses graphical elements to convey programming concepts, allowing individuals to develop software by manipulating visual components instead of writing code~\cite{myers1990taxonomies}. This graphical representation is often more intuitive and accessible than traditional programming languages, lowering the barriers to entry for beginners or non-programmers in software development~\cite{whitley1997visual, myers1986visual}. Visual programming systems are generally categorized into two main types: flowchart-based and block-based.

Flowchart-based visual programming represents programming logic using flowchart diagrams, with each node symbolizing a distinct operation or control flow~\cite{cox1989prograph, kodosky2020labview, chen2021entanglevr, hooshyar2015flowchart, yigitbas2023end}. This method typically involves using functions or APIs as discrete units and connecting these units as nodes in a network, making it particularly effective for visually illustrating complex logic and control structures. However, managing complex programs can become challenging due to screen space limitations and the potential for an overwhelming number of node connections.

Block-based visual programming represents programming logic using interlocking blocks, each corresponding to a specific operation or data type~\cite{ye2024prointerar, lin2024jigsaw, resnick2009scratch, blockly, jung2021blocklyxr, kelly2018arcadia, appinventor, zhu2023learniotvr}. Users can perform drag-and-drop actions on these blocks to build programs, linking them to form sequences of operations. This approach offers greater flexibility than flowchart-based visual programming, as it allows for nesting and reuse of code blocks. Nonetheless, visualizing complex logic can still be challenging, as most elements are closely tied to programming concepts.

This study introduces \sysname{}, a tool that addresses the limitations of traditional visual programming by leveraging LLMs to abstract away programming complexities. Using a simplified flowchart approach combined with natural language prompts, \sysname{} enables non-technical users to create AI services.

\subsection{Integration of LLMs in Visual Programming}
The emergence of LLMs has brought significant advancements to the field of visual programming. Several empirical studies have examined LLMs’ potential as development tools~\cite{finnie2022robots, hellas2023exploring, leinonen2023comparing, sarsa2022automatic}, and comparisons between LLM-generated and human-authored code and explanations~\cite{hellas2023exploring, leinonen2023comparing, perry2023users} have demonstrated that, when leveraged properly, LLMs can provide substantial value to developers and students. These models possess the ability to generate human-like text and understand user intent~\cite{nam2024using}, enabling them to offer context-appropriate suggestions and modifications. This contributes to overcoming limitations in visual programming that were previously difficult to address due to their code-centric nature. Unlike traditional visual programming approaches that substitute functions or variables with visual elements, integrating LLMs allows user commands and intentions to directly serve as inputs, thereby enhancing both the intuitiveness and efficiency of visual programming.

Recent human-computer interaction (HCI) research has actively explored the integration of LLMs with visual programming. 
Notable examples include PromptMaker~\cite{jiang2022promptmaker}, which enables AI/ML model prototyping through natural language prompts; Rapsai~\cite{du2023rapsai}, which supports rapid prototyping of AI-based multimedia applications; and SEAM-EZ~\cite{yu2024seam}, which simplifies stateful analytics through visual programming. Furthermore, research on building LLM pipelines using visual programming continues to expand~\cite{wu2022promptchainer, comfyui, flowiseai, langflow, wu2022ai, cheng2024prompt, zhou2023instructpipe, carney2020teachable, wei2022chain}. PromptChainer~\cite{wu2022promptchainer} provides a visual programming chain interface for building various application prototypes, while Low-code LLM~\cite{cai2023low} offers an environment where users can input task prompts, collaborate with LLMs to decompose tasks, and generate and execute workflows. AI Chain~\cite{wu2022ai} proposes an interactive system applying the chaining concept, where the output of one step becomes the input for the next. This approach not only improves task outcome quality but also significantly enhances system transparency, controllability, and collaborative experience.

\subsection{User Cognitive Challenges in Natural Language LLM Interfaces}
Natural language interactions with LLMs are transforming HCI paradigms while simultaneously presenting users with unique cognitive challenges. The research by Subramonyam et al.~\cite{subramonyam2024bridging} and Zamfirescu-Pereira et al.\cite{johnny} comprehensively examines the user problems that emerge in these new interaction modes and their underlying causes.

Subramonyam et al.~\cite{subramonyam2024bridging} introduced the concept of a gulf of envisioning to characterize the unique challenges of interacting with LLMs. This framework consists of three sub-gaps:
(1) the capability gap, where users lack clear mental models of what LLMs can or cannot do, impeding proper expectation setting;
(2) the instruction gap, which reflects the difficulty of translating user intentions into prompts that LLMs can reliably interpret; and
(3) the intentionality gap, representing the absence of cognitive criteria for evaluating outputs and adjusting them toward desired outcomes.

The research by Zamfirescu-Pereira et al.\cite{johnny} provides concrete behavioral evidence supporting this theoretical framework. By observing 10 non-experts designing LLM prompts, the researchers discovered the following patterns: (1) reliance on opportunistic trial-and-error instead of systematic approaches, (2) hasty generalization from single instances, (3) inappropriate application of human-human interaction norms to LLMs, (4) preference for direct instructions over providing examples, and (5) focus on limited scenarios rather than systematic testing. These patterns can be interpreted as practical manifestations of the design gaps identified by Subramonyam et al.\cite{subramonyam2024bridging}.

The integrated perspective from both studies offers important insights into the unique characteristics of LLM interfaces and the resulting user challenges. Unlike traditional interfaces with clearly defined functions, LLM-based interactions allow users to express diverse intentions through natural language and generate probabilistic outputs. These characteristics offer users a high degree of freedom and flexibility, but also increase the cognitive burden required for effective interaction.

For effective LLM interface design, approaches such as the six design patterns proposed by Subramonyam et al.\cite{subramonyam2024bridging}—visual tracking of prompts and outputs, providing ideas for prompt writing, offering multiple outputs, making results explainable, using domain-specific prompting strategies, and allowing manual control of outputs—are necessary. Zamfirescu-Pereira et al.\cite{johnny}'s research complements this by emphasizing the need for systematic prompt testing mechanisms, education about effective prompting strategies, and interfaces that clarify the special nature of human-machine interactions.

In addition to these challenges, recent work has underscored the role of query reformulation in mitigating cognitive burdens. Several approaches have explored the use of intelligent agents to enhance both prompt refinement and workflow orchestration. Specifically, query expansion and planning can be realized through dedicated agent-based methods. Query expansion utilizes LLM-based techniques to analyze the initial user prompt, broadening the input by incorporating relevant synonyms, related concepts and contextual cues~\cite{carpineto2012survey,wang2023query2doc,jagerman2023query}. This process effectively clarifies ambiguous or complex queries, ensuring that even incomplete instructions are transformed into precise, context-aware commands. Meanwhile, planning agents organize these enriched inputs into a coherent sequence of actionable steps by mapping dependencies between actions and refining the overall workflow for optimal execution~\cite{wang2023describe,li2024autoflow,huang2024planning}. Such agent-based methods, combined with traditional approaches, enable systems better manage the context of complex tasks and ultimately deliver more robust, also efficient service execution.

\begin{figure*}[!ht]
      \centering
      \includegraphics[width=1.00\textwidth]{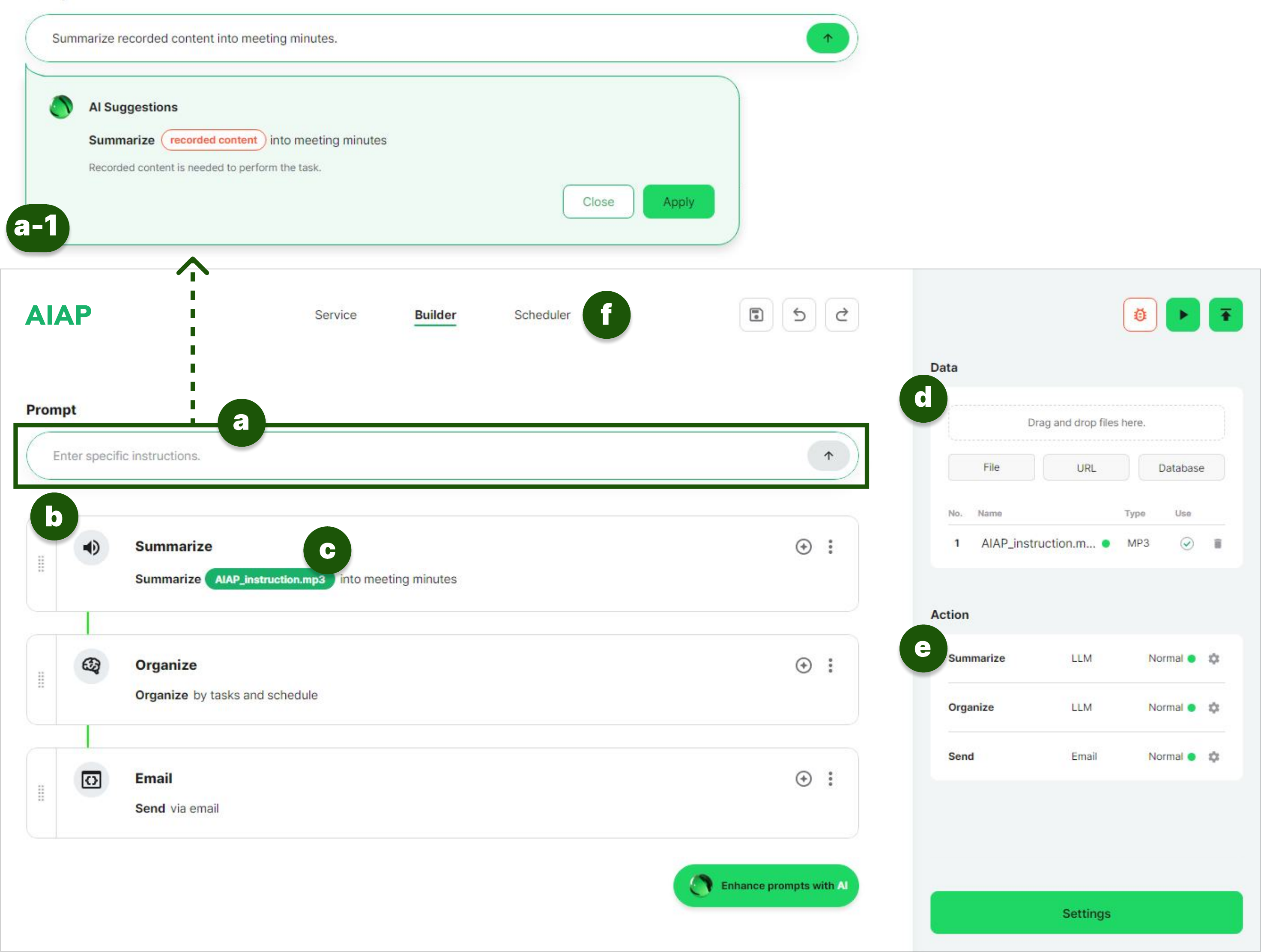}
      \caption{A builder page for creating AI services with an example screen for Task 1 in a study 1: (a) A section for entering desired instructions. Once input is provided, an AI-Generated Suggestion appears to interpret the prompt (a-1). Since data definition is required, it is labeled as \redbox{recorded content}. (b) A unidirectional modular step system where user inputs are accumulated step-by-step, and the service is executed in sequence. (c) The interpreted prompt. Data connections are displayed as capsules, with the action section highlighted in bold. When data is connected, it appears as \greenbox{AIAP\_instruction.mp3}. (d) The data field, which allows for the registration of files, URLs, and databases, and connects them to the prompt. (e) The action field, which automatically identifies actions from the prompt and displays related functionalities or APIs for automatic linking. (f) A menu to switch between the Service tab and the Schedule tab. (Task 1 of the comparative evaluation}
      \Description{Builder page for AI service creation: a modular system with user input-based interpreted prompts, data connections, step-by-step service execution, file/URL/database registration, and automated function linking.}
      \label{fig:aiap1}
\end{figure*}

\section{System Design}
Designing AI systems for non-experts requires both a clear understanding of user intentions and effective mechanisms for translating them into actionable results. \sysname{} supports this transformation by turning users’ creative intentions into executable AI workflows. Our goal is to provide an environment where non-experts can intuitively explore the potential of AI systems, transform abstract ideas into concrete implementations, and design effective workflows—all without requiring complex coding knowledge.

\sysname{} is designed based on the principle that "intentions are actions," as proposed by Subramonyam et al.\cite{subramonyam2024bridging}. According to this research, the effective connection between user intentions and system behaviors is a key element of successful AI interactions. To achieve this, our system is specifically designed to address three major gaps in AI interaction: the \textit{intentionality gap} stemming from difficulties in expressing user goals, the \textit{capability gap} arising from misunderstandings about AI capabilities, and the \textit{instruction gap} involving challenges in translating intentions into effective prompts. Building on this design philosophy, \sysname{} integrates four core functions: (a) AI-powered suggestions, (b) Modular workflow management, (c) Automatic identification of data, actions, and context, and (d) Intuitive modification with intelligent connections.

\subsection{AI-Generated Suggestion}
As shown in Figure~\ref{fig:aiap1} (a-1), the AI-Generated Suggestion feature of \sysname{} interprets and restructures user requirements into coherent, actionable steps. These suggestions are presented for user confirmation before the workflow is composed. This process functions similarly to an auto-complete mechanism, assisting users in concretizing their goals.
This feature is designed to mitigate the \textit{intentionality gap} and \textit{instruction gap}~\cite{subramonyam2024bridging}, which arises when users struggle to articulate their intentions in a structured, specific, and sequentially appropriate manner for execution.

To address this challenge, the AI-Generated Suggestion module systematically analyzes fragmented and unordered user inputs, reconstructing them into clear and executable steps.  
Users can review and approve these structured suggestions before proceeding to the visual workflow construction phase.  
By adding structure to the opportunistic and ad-hoc prompt patterns commonly observed among non-experts~\cite{johnny}, this feature reduces cognitive load, prevents common prompting errors, and helps users develop more sophisticated AI services with less trial and error. The result is a development process that is both more intuitive for newcomers and more efficient for experienced users.

\subsection{Modular Workflow Management with Nodes}
Workflows in \sysname{} are organized into modular Nodes, enabling users to easily manage and adjust individual components. This modularity supports intuitive drag-and-drop interactions, allowing users to quickly rearrange or modify workflow Nodes. Such modularization significantly improves readability, facilitates maintenance, and ensures workflow flexibility. (See figure~\ref{fig:aiap1} (b)).

\subsection{Automatic Identification of Data, Action, and Context}
\sysname{} analyzes user-inputted natural language instructions and automatically identifies key elements based on the linguistic structure of the sentence. Specifically, nouns are categorized as \greenbox{Data}, verbs are displayed in \textbf{bold as Actions}, and additional descriptive phrases are classified as \textit{Context}. These components are visually highlighted within the interface, allowing users to immediately understand how their instructions are interpreted—without the need for any technical knowledge. 

In particular, for Data elements, \sysname{} visually distinguishes between unconnected and connected inputs. Data that is not yet linked to any file or source is labeled as \redbox{Data}, whereas successfully connected items are indicated with \greenbox{Data}. This distinction helps users quickly recognize which elements still require input and which have already been resolved.  (See figure~\ref{fig:aiap1} (c)).

This process allows users to intuitively grasp the structure of their workflow and clearly see how data sources, actions, and context are connected—enhancing both the transparency and efficiency of the workflow design process.

\subsection{Intelligent Action Linking}
The Intelligent Action Linking feature of \sysname{} automatically maps user-described actions to appropriate LLMs, tools, or APIs. By doing so, it addresses the \textit{capability gap}~\cite{subramonyam2024bridging}—bridging users' limited understanding of what the system can or cannot do.

Rather than requiring users to manually select tools or understand backend functionalities, the system infers the necessary services based on natural language input and seamlessly links them to the intended actions.  
This automated mapping not only streamlines the development process but also helps users develop a clearer understanding of the system’s capabilities through transparent and contextual suggestions.  (See figure~\ref{fig:aiap1} (e).

By abstracting away the complexity of tool selection and integration, \sysname{} improves the accessibility and usability of AI technologies, especially for non-expert users.

\section{Methodology: Multi-Agent Collaboration Behind \sysname{}}
Multi-agent collaboration is essential to \sysname{}'s functionality. The complex process of translating non-experts' ambiguous natural language requests into functional AI workflows requires specialized expertise across multiple domains. A single agent would struggle to achieve high accuracy while simultaneously interpreting user intentions, identifying appropriate tools, establishing API connections, and optimizing workflows. By distributing these responsibilities among specialized agents that each excel in different aspects of the development process, \sysname{} achieves greater reliability, adaptability, and precision while maintaining a simple user-facing interface that effectively bridges the identified \textit{capability}, \textit{instruction}, and \textit{intentionality gaps}.

In this section, we detail our proposed framework for processing user queries, assigning plans, and executing actions. It consists of four main modules: 
(1) \emph{Query Process Module}, 
(2) \emph{Task Planning \& Entity Extraction Module}, 
(3) \emph{Action Mapping \& Execution Module}, 
and (4) \emph{Plan Refinement Module (Human in the Loop)}.

\subsection{Query Process Module}
Given an initial user query \(Q\), the \emph{Query Process Agent} first determines which operation is most appropriate based on the features of \(Q\). Specifically, the agent selects one of three operations: query reformulation, expansion, or decomposition. This decision is modeled as:
\begin{equation}\label{eq:query-process}
Q' = G_\mathcal{Q}(Q, \text{option}),
\end{equation}
where \(\text{option} \in \{\text{reformulation}, \text{expansion}, \text{decomposition}\}\) and \(G_\mathcal{Q}\) represents the query processing function. The output \(Q'\) is a revised query that is precise and contextually enriched, thereby laying the groundwork for accurate task planning.

\subsection{Task Planning \& Entity Extraction Module}
Once the refined query \(Q'\) is produced, it is passed to the \emph{Task Planning Agent}, which decomposes the query into a sequence of actionable steps:
\begin{equation}\label{eq:task-planning}
S = \{p_1, p_2, \dots, p_k\} = G_\mathcal{P}(Q'),
\end{equation}
where \(G_\mathcal{P}\) denotes the planning function and each \(p_i\) corresponds to a discrete action or subtask. For instance, if the user asks, “Please search for a specific book on Google and then buy it,” the planning agent may decompose this into two steps: \(\texttt{searchBook}\) and \(\texttt{purchaseBook}\).
In parallel, the \emph{Entity Extraction Agent} identifies key entities and attributes from the planned steps. This process tags relevant details (e.g., \emph{book title}, \emph{search engine}, \emph{purchase platform}) that are needed for subsequent modules:
\begin{equation}\label{eq:entity-extraction}
E = G_\mathcal{E}(S),
\end{equation}
where \(G_\mathcal{E}\) is the entity extraction function, and \(E\) is the set of extracted entities associated with each planned action.

\subsection{Action Mapping \& Execution Module}
The third module, \emph{Action Mapping \& Execution}, bridges the plan \(S\) to actual implementations. We maintain an \emph{Action Pool} \(\mathcal{A}\) containing a set of predefined actions (e.g., APIs or services). Each action \(\alpha \in \mathcal{A}\) is represented by an embedding that encodes its functionality. For each step \(p_i \in S\), the system first retrieves the top \(k\) candidate actions \(\mathcal{C}\) from \(\mathcal{A}\) based on similarity scores computed between \(\mathrm{embed}(p_i)\) and \(\mathrm{embed}(\alpha)\) for each \(\alpha \in \mathcal{A}\):
\begin{equation}\label{eq:action-retrieval}
\mathcal{C} = \text{Top}_k\Bigl\{ \alpha \in \mathcal{A} : \mathrm{sim}\bigl(\mathrm{embed}(p_i), \mathrm{embed}(\alpha)\bigr) \Bigr\}.
\end{equation}
Subsequently, a dedicated \emph{Mapping Agent} selects the most appropriate candidate from the retrieved pool:
\begin{equation}\label{eq:action-mapping}
\alpha^* = \arg\max_{\alpha \in \mathcal{C}} R(\alpha),
\end{equation}
where \(R(\alpha)\) denotes the mapping agent's score for action \(\alpha\). Once the best-matching action is identified, it is executed with the appropriate parameters extracted from \(E\). This process ensures that the plan is both context-aware and precisely mapped to the correct system functionality.

\subsection{Plan Refinement Module (Human in the Loop)}
After the action mapping and execution, the system incorporates a human-in-the-loop phase via the \emph{Plan Refinement Module}. In this phase, the executed actions and the corresponding plan are presented to a human operator (or via a chatbot interface) for evaluation. The operator can either approve the executed plan or provide feedback for modification. If the plan is unsatisfactory, the human feedback \(f_{\text{human}}\) is used to adjust the plan. This iterative refinement process is modeled as:
\begin{equation}\label{eq:plan-refinement}
S^{(n+1)} = G_\mathcal{P}\Bigl(S^{(n)}, f_{\text{human}}^{(n)}\Bigr), \quad n = 0, 1, 2, \dots, N,
\end{equation}
with \(S^{(0)}\) being the initial plan prior to execution and \(S^{*} = S^{(N)}\) being the final, approved plan. This feedback loop ensures that any misalignment between intended and executed actions is identified and corrected, leading to an optimal and robust execution strategy aligned with user goals.

\subsection{Technical Implementation}

The multi-agent system behind \sysname{} is implemented using a modular and production-ready technology stack that ensures reliable operation and efficient interaction among agents.  The user interface is built using $Next.js$ and $React$, ensuring an intuitive experience. Natural language processing and agent interactions are powered by $GPT-4o$ (via OpenAI API) integrated with $LangChain$\footnote{\url{https://www.langchain.com}}, which efficiently manages prompt engineering and AI-agent orchestration. Real-time synchronization and feedback mechanisms are enabled through $Server-Sent Events (SSE)$, ensuring instant communication and responsiveness. Additionally, backend workflow management employs $Zod Schemas$, $MySQL$, and $Drizzle ORM$, chosen for their reliability and scalable data handling capabilities. This technology stack ensures seamless real-time processing and robust interactions within the \sysname{} multi-agent environment. Moreover, for the retrieval task, we employed the \emph{Multilingual-E5-base} model~\cite{wang2024multilingual}\footnote{\url{https://huggingface.co/intfloat/multilingual-e5-base}}, which generates embeddings and retrieves the top $10$ most relevant APIs based on cosine similarity.

\begin{figure*}[!ht]
      \centering
      \includegraphics[width=1.00\textwidth]{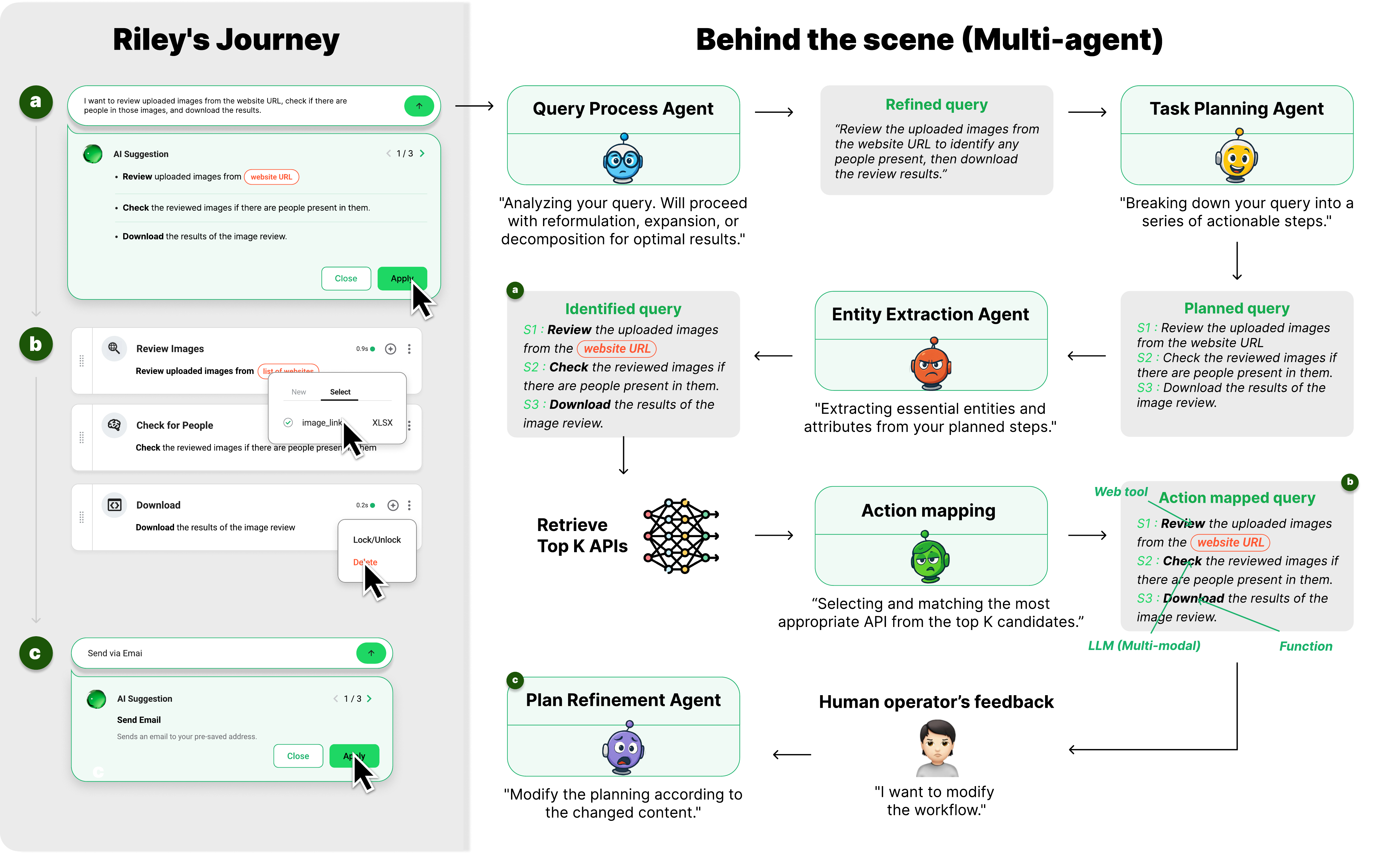}
      \caption{Overview of AIAP’s multi-agent orchestration. On the left, Riley’s Journey illustrates the user-facing steps for creating AI workflows, including entering natural language prompts, reviewing data, and arranging modular actions. On the right, specialized multi-agents collaboratively handle query processing, entity extraction, api retrieval, action mapping, and plan refinement, automatically transforming user instructions into structured, executable workflows.}
      \Description{Overview of AIAP’s multi-agent orchestration. On the left, Riley’s Journey illustrates the user-facing steps for creating AI workflows, including entering natural language prompts, reviewing data, and arranging modular actions. On the right, specialized multi-agents collaboratively handle query processing, entity extraction, api retrieval, action mapping, and plan refinement, automatically transforming user instructions into structured, executable workflows.}
      \label{fig:users}
\end{figure*}

\section{User Scenario: Riley's Journey with \sysname{}}
To illustrate how \sysname{} operates in practice, we present a user scenario featuring a fictional character, Riley. Although the name subtly references the protagonist of Inside Out~\cite{pixar2015insideout}, here Riley serves as a non-expert user navigating a real-world challenge. Similar to how Riley in Inside Out is influenced by hidden agents controlling her emotions, our fictional Riley’s interactions with \sysname{} are powered by a team of intelligent agents operating behind the scenes—each responsible for interpreting language, planning tasks, and linking tools. Each agent contributes a distinct function—such as language interpretation, task planning, or tool integration—within a collaborative process.

Riley is a security manager at a major e-commerce company, responsible for monitoring thousands of user-uploaded product review images daily to ensure that no personal information appears in the content—a violation of platform policy. With limited technical background, she found existing automation tools difficult to use and turned to \sysname{} for a more intuitive solution.

This narrative highlights both the simplicity of the user-facing interface and the sophisticated orchestration of the system’s core functionalities: AI-generated suggestions, modular workflow management, intelligent action linking, and contextual data handling—all made possible through multi-agent collaboration.

\subsection{Requesting a Task}
Riley begins by entering a prompt:  
“I want to review uploaded images from the website, check if there are people in those images, and download the results.” (See Figure~\ref{fig:users} (a)).

While the request may appear simple on the surface, it triggers a coordinated sequence of actions among specialized agents within \sysname{}’s architecture.
The \emph{Query Process Agent} analyzes the input, deciding whether it should be reformulated or decomposed.  
Next, the \emph{Task Planning Agent} breaks the query into manageable steps.  
The \emph{Entity Extraction Agent} then labels the nouns, verbs, and descriptors to form structured components: data, actions, and context.

Their collaborative effort results in a structured, human-readable action list presented to Riley:

\begin{enumerate}[label=(\arabic*)]
    \item \textbf{Review} uploaded images from \redbox{website URL}
    \item \textbf{Check} the reviewed images if there are people present in them.
    \item \textbf{Download} the results of the image review.
\end{enumerate}

Through the \textit{AI-Generated Suggestion} feature, Riley immediately sees that the system understands her intent.  
She clicks \textit{Apply} to generate the corresponding workflow—without needing to write a single line of code or understand how the query was processed.

\subsection{Adding Required Information}
The visual workflow appears: a series of nodes connected in sequence.  
This represents \sysname{}’s \emph{Modular Workflow Management} interface, allowing Riley to directly interact with each component through drag-and-drop.

She notices that the first node contains a red \redbox{website URL}, indicating that required input is missing.  (See Figure~\ref{fig:users} (b)).
Responding to this cue, Riley drags her Excel file (\texttt{image\_link.xlsx}) into the node.  
The placeholder instantly changes to \greenbox{image\_link.xlsx}, confirming the input.

Meanwhile, hidden from view, the \emph{Mapping Agent} is activated.  
It evaluates the planned task, retrieves candidate APIs for image analysis, and selects the best fit—automatically binding it to the node.

Riley doesn't need to think about API endpoints or parameter formats. The system visually confirms that her input has been processed correctly, allowing Riley to recognize progress without needing to understand the underlying mechanics.

\subsection{Modifying the Workflow}
Initially, the workflow ends with a node to download the results.  
But Riley realizes that receiving the results via email would be more practical for her daily routine.

She removes the “Download” node and adds a “Send via Email” node in its place using the visual interface. (See Figure~\ref{fig:users} (c)).
This action triggers the \emph{Plan Refinement Agent}, which checks that the new configuration is logically valid and automatically updates dependencies.

Riley watches as the system adjusts itself instantly. The workflow remains intact, and she feels reassured that even manual edits won’t break the process.

\subsection{Completing and Automating the Workflow}
Riley clicks the “Run” button to execute the workflow.  
Each node activates in order, and the interface provides live feedback.  
When it gets completed, a message confirms that the results have been sent to her email.

\begin{figure}
      \centering
      \includegraphics[width=0.5\textwidth]{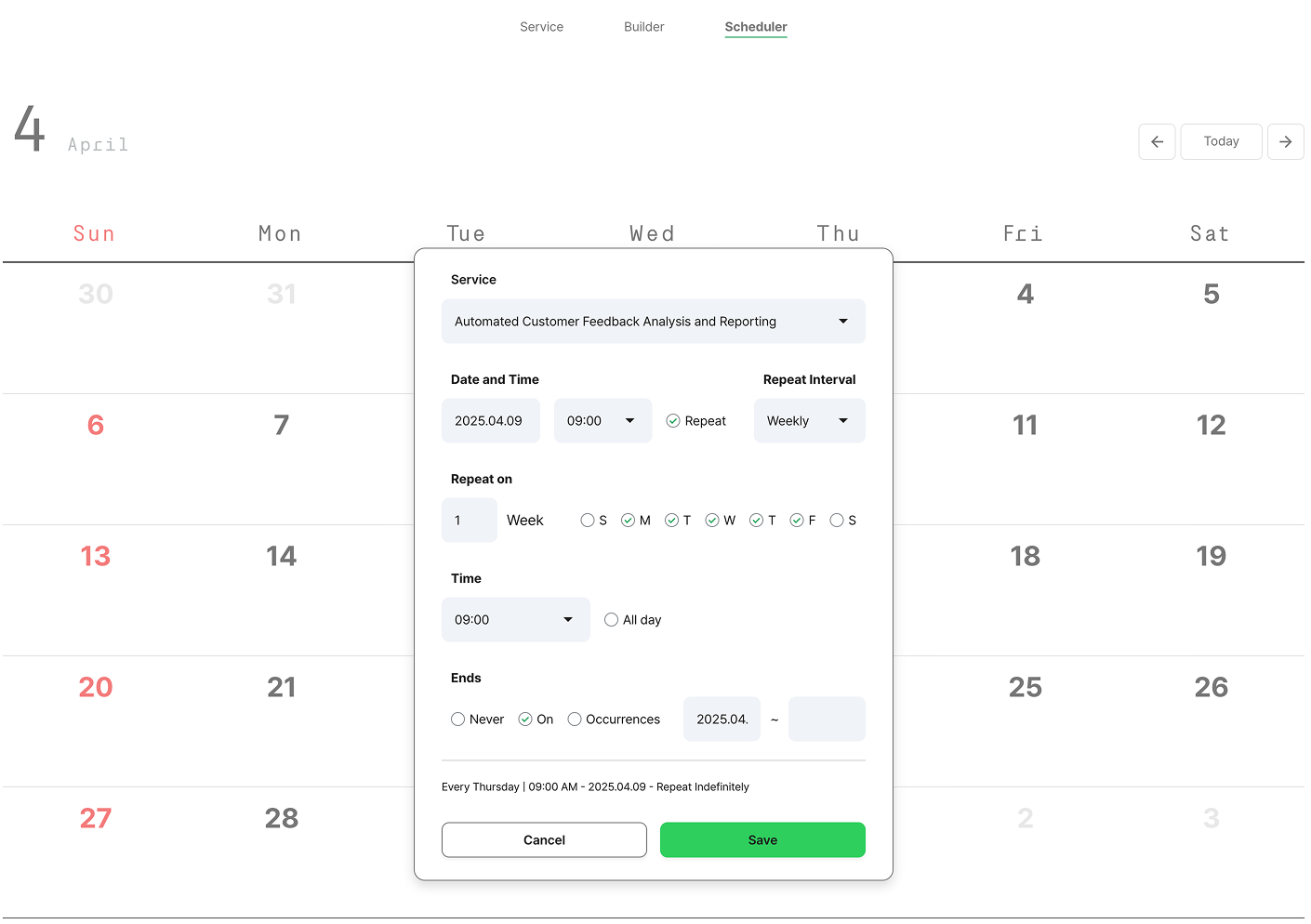}
      \caption{Average NASA-TLX Item Scores by Task. Lower scores indicate better outcomes (i.e., lower workload). Performance was reversed to align directionality.}
    \Description{Average NASA-TLX Item Scores by Task. Lower scores indicate better outcomes (i.e., lower workload). Performance was reversed to align directionality.}
      \label{fig:schedule}
\end{figure}

She then opens the scheduler tab and configures the workflow to run automatically every day at 9:00 AM.  
A routine manual task has been successfully automated with minimal user effort. (See Figure~\ref{fig:schedule}).

Throughout this experience, Riley never had to worry about system internals.  
Her role was simply to express intent. The agents inside \sysname{}—like the characters inside her mind—handled the complexity for her.  
AI, once perceived as complicated and inaccessible, now feels approachable, responsive, and aligned with her thinking.

% \begin{table*}
%   \caption{Some Typical Commands}
%   \label{tab:commands}
%   \begin{tabular}{ccl}
%     \toprule
%     Command &A Number & Comments\\
%     \midrule
%     \texttt{{\char'134}author} & 100& Author \\
%     \texttt{{\char'134}table}& 300 & For tables\\
%     \texttt{{\char'134}table*}& 400& For wider tables\\
%     \bottomrule
%   \end{tabular}
% \end{table*}

\begin{table*}
  \caption{User study participants: employees who develop and utilize basic AI services, and students who integrate AI into their
academic work.}
  \Description{User study participants: employees who develop and utilize basic AI services, and students who integrate AI into their
academic work.}
  \label{tab:participants}
\begin{tabular}{cccll}
    \toprule
        \textbf{Study} & \textbf{Participant} & \textbf{Age} & \textbf{Role}                  & \textbf{AI Proficiency (5-point Scale)} \\
    \midrule
        1                & P1                   & 30s          & Employee (Business developer)  & 2 (Basic AI concepts)                   \\
        1                & P2                   & 30s          & Employee (Front-end developer) & 3 (Explain simple AI model)             \\
        1                & P3                   & 30s          & Employee (Product owner)       & 3 (Explain simple AI model)             \\
        1                & P4                   & 20s          & Graduate student               & 2 (Basic AI concepts)                   \\
        1                & P5                   & 20s          & Graduate student               & 3 (Explain simple AI model)             \\
        1                & P6                   & 20s          & Graduate student               & 3 (Explain simple AI model)             \\
        1                & P7                   & 20s          & Graduate student               & 2 (Basic AI concepts)                   \\
        1                & P8                   & 20s          & Graduate student               & 3 (Explain simple AI model)             \\
        1                & P9                   & 20s          & Graduate student               & 2 (Basic AI concepts)                   \\
        1                & P10                  & 20s          & Graduate student               & 3 (Explain simple AI model)             \\
        1                & P11                  & 20s          & Graduate student               & 2 (Basic AI concepts)                   \\
        1                & P12                  & 20s          & Employee (Operation)           & 2 (Basic AI concepts)                   \\
        1                & P13                  & 20s          & Graduate student               & 2 (Basic AI concepts)                   \\
        1                & P14                  & 30s          & Employee (Backend developer)   & 2 (Basic AI concepts)                   \\
        1                & P15                  & 30s          & Employee (Product owner)       & 2 (Basic AI concepts)                   \\
        1                & P16                  & 30s          & Employee (Business support)    & 3 (Explain simple AI model)             \\
        1                & P17                  & 30s          & Employee (Business support)    & 2 (Basic AI concepts)                   \\
        1                & P18                  & 20s          & Employee (Business support)    & 2 (Basic AI concepts)                   \\
        1                & P19                  & 30s          & Employee (Backend developer)   & 3 (Explain simple AI model)             \\
        1                & P20                  & 30s          & Employee (Product manager)     & 2 (Basic AI concepts)                   \\
        1                & P21                  & 30s          & Employee (Product owner)       & 3 (Explain simple AI model)             \\
        1                & P22                  & 20s          & Graduate student               & 2 (Basic AI concepts)                   \\
    \midrule
        2                & P23                  & 20s          & Graduate student               & 3 (Explain simple AI model)             \\
        2                & P24                  & 20s          & Graduate student               & 2 (Basic AI concepts)                   \\
        2                & P25                  & 20s          & Graduate student               & 2 (Basic AI concepts)                   \\
        2                & P26                  & 20s          & Graduate student               & 2 (Basic AI concepts)                   \\
        2                & P27                  & 20s          & Graduate student               & 2 (Basic AI concepts)                   \\
        2                & P28                  & 30s          & Employee (QA tester)           & 2 (Basic AI concepts)                   \\
        2                & P29                  & 30s          & Employee (Business support)    & 2 (Basic AI concepts)                   \\
        2                & P30                  & 30s          & Employee (Product manager)     & 2 (Basic AI concepts)                   \\
        2                & P31                  & 30s          & Employee (Business developer)  & 2 (Basic AI concepts)                   \\
        2                & P32                  & 30s          & Employee (Product owner)       & 2 (Basic AI concepts)                  \\
    \bottomrule
\end{tabular}
\end{table*}

\section{User Study}
To progressively assess the usability and user experience of \sysname{}, we conducted a two-stage user study. The first stage, a basic usability study, was aimed at evaluating whether users could successfully complete predefined tasks with the help of multi-agents and features of \sysname{} and to identify potential areas for improvement. Based on the findings from this study, we refined \sysname{} to enhance its usability and enable users to construct a wider range of workflows. The second stage, a free-exploration user experience study, aimed to capture more authentic user behavior, providing insights not only into usability but also broader experiential factors in practice where multi-agents can support.

\subsection{Study 1: Basic Usability Study}
In the initial stage, we conducted a controlled experiment to evaluate the fundamental usability of \sysname{} by asking participants to complete a set of predefined tasks.

\subsubsection{Participants}
Participants were recruited based on their basic familiarity with AI and recognition of its relevance to their professional or academic work. Specifically, we targeted two groups: (1) professionals who regularly need to create or use basic AI-driven services in their workplace, and (2) university students and graduate researchers who aim to integrate AI services into their studies or research. A total of 22 participants were recruited through both announcements and snowball sampling. The group consisted of professionals and graduate students, with an average age of 30.45 years ($SD$: 4.09), including 7 females and 15 males. Participants who had no prior experience or interest in AI/ML, or those with extensive expertise as AI/ML developers or data scientists, were excluded to maintain a focus on non-expert users. On a 5-point Likert scale, the average self-reported familiarity with AI was 2.45. Most participants had limited prior experience with LLM-based tools such as ChatGPT~\cite{chatgpt}. Each participant was compensated at a rate of \$20 per hour.

\subsubsection{Procedure}
The study was conducted individually, either in person or via Zoom, depending on participant preference. Each session lasted approximately 1 hour and 20 minutes, during which both audio and screen activity were recorded. After obtaining informed consent, participants received a brief introduction to the experimental protocol and a demonstration of the \sysname{} interface. Participants were then given time to independently explore the system.

Following this exploration phase, the researcher demonstrated each of the three tasks using \sysname{}, after which participants independently performed the same tasks. After each task, participants completed a questionnaire to evaluate workload and usability using the NASA-TLX~\cite{hart1988development} and SUS~\cite{bangor2008empirical} scales—both widely adopted instruments for system usability assessment.

After completing all tasks, participants engaged in a semi-structured follow-up interview. During the interview, they revisited each task to provide in-depth feedback on their experience, including reflections on specific steps and features within \sysname{}.

\subsubsection{Tasks}
We designed three tasks, considering realistic, practical use cases, where multi-agents mechanisms and AI features of the tool can support. These tasks were chosen to assess \sysname{}'s effectiveness across both routine and specialized workflow scenarios.

\paragraph{Task 1. Compose a summary of the recorded meeting minutes and send via email.}
This task simulated the repetitive documentation work often performed by office professionals. Participants entered the prompt \textit{"Summarize recorded content into meeting minutes"}, linked a recorded file using the \redbox{recorded content} capsule, and then used the prompts \textit{"Organize by tasks and schedule"} and \textit{"Send via email"}.
\sysname{} successfully processed and structured the content before sending the result via email. Although the task benefited from multi-agent collaboration—from query to execution—the entire process was handled automatically in the background, while users remained unaware of the underlying orchestration performed by the multi-agent system.

\paragraph{Task 2. Structure the paper into bullet points, translate the content, incorporate references, and prepare it for download.}
This task involved automating time-consuming research-related processes. Participants entered the prompt \textit{"Organize the paper into bullet points"} and linked a file using the \redbox{paper} capsule. They then used prompts such as \textit{"Translate to Korean"}, \textit{"Add additional reference materials"}, and \textit{"Download"}, which enabled \sysname{} to generate a summarized, translated version of the paper with references, ready for download.

\paragraph{Task 3. Examine the provided list of image URLs and indicate with an O or X whether they depict human subjects.}
This task represented specialized, high-demand workflows. Participants used the prompt \textit{"Indicate O if there is a person and X if there is not on list website URL"} and uploaded a file of image URLs via the \redbox{website URL} capsule. They then used the prompt \textit{"Send via email"} and scheduled the task to run automatically every Wednesday at 9:00 a.m.
%User scenario와 동일하다는 내용?.

\begin{figure}
      \centering
      \includegraphics[width=0.5\textwidth]{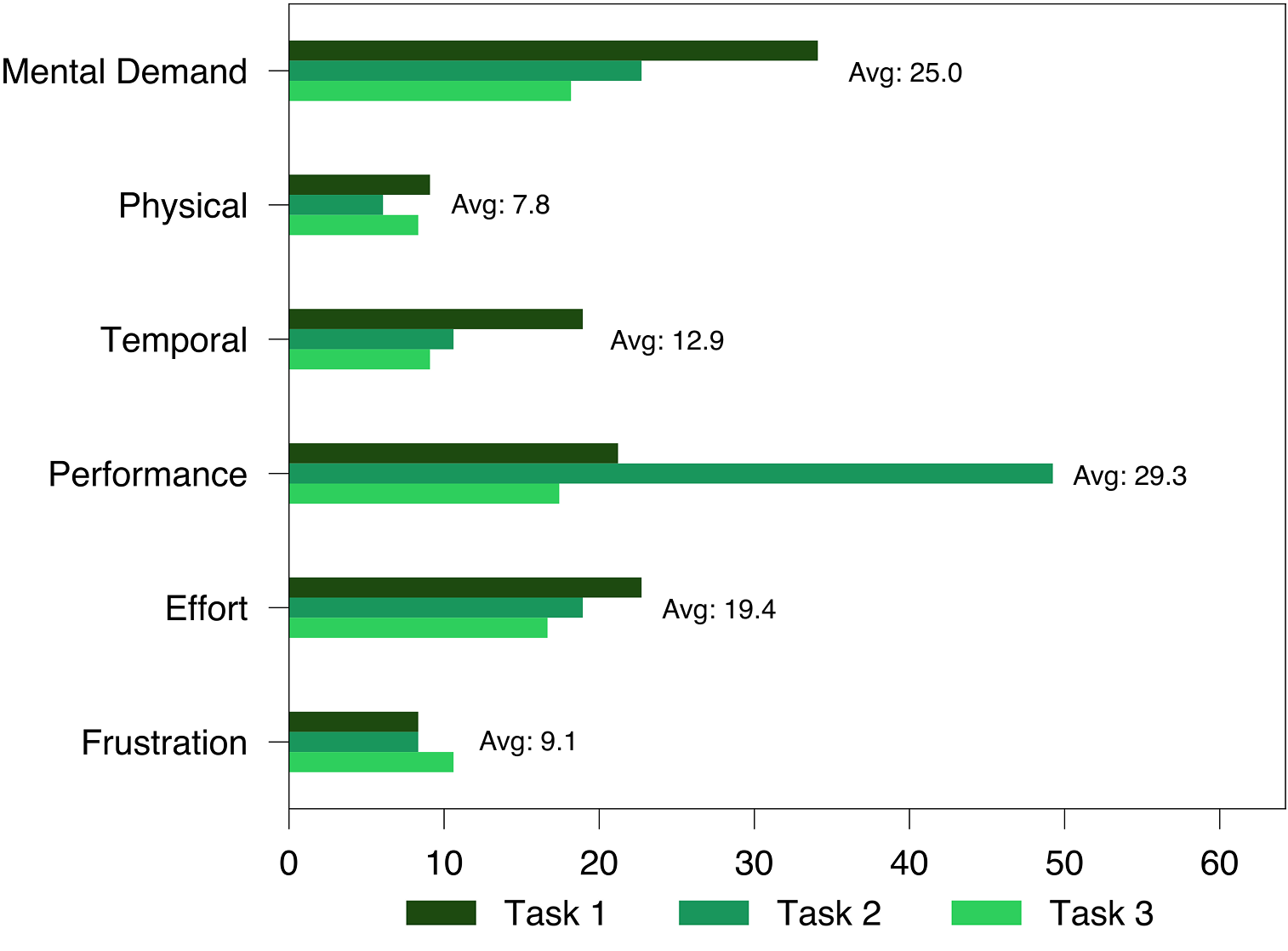}
      \caption{Average NASA-TLX Item Scores by Task. Lower scores indicate better outcomes (i.e., lower workload). Performance was reversed to align directionality.}
    \Description{Average NASA-TLX Item Scores by Task. Lower scores indicate better outcomes (i.e., lower workload). Performance was reversed to align directionality.}
      \label{fig:nasatlx}
\end{figure}

\begin{figure}
      \centering
      \includegraphics[width=0.5\textwidth]{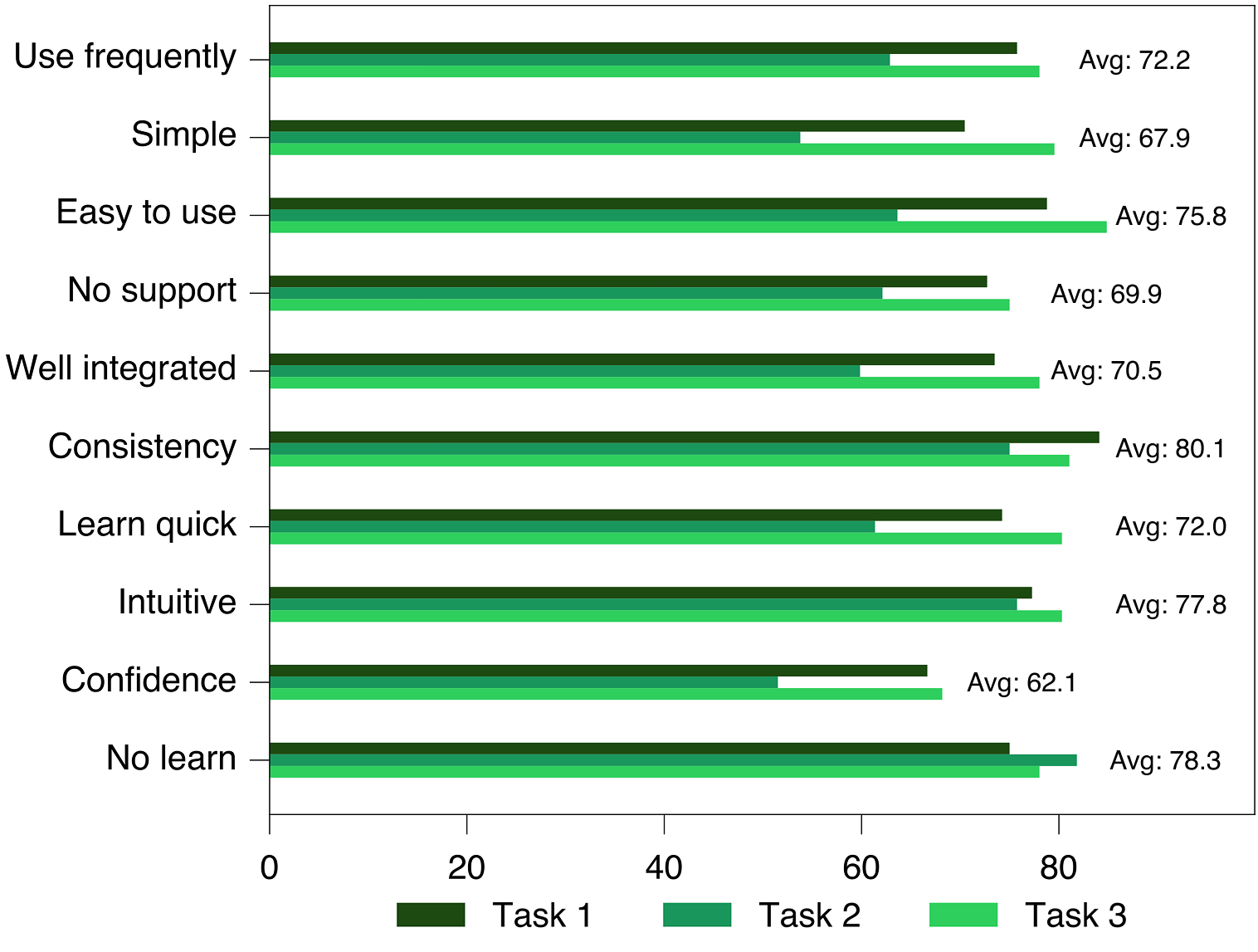}
        \caption{Average SUS Item Scores by Task. Task-wise Average Scores for Each SUS Item. All scores range from 0 to 100, with higher scores indicating better usability.}
        \Description{Average SUS Item Scores by Task. Task-wise Average Scores for Each SUS Item. All scores range from 0 to 100, with higher scores indicating better usability.}
      \label{fig:sus}
\end{figure}

\subsubsection{Results}
First, according to the NASA-TLX data, the overall average workload score was 17.26, indicating that \sysname{} provided a generally satisfactory user experience. When broken down by task, Task 1 scored 19.07, Task 2 scored 19.32, and Task 3 scored 13.38. While there were slight variations between tasks, the results indicate consistent usability across diverse workflow scenarios, with Task 3 showing notably lower workload. A closer look at specific subscales showed that scores for Mental Demand (M=25.0) and Performance (M=29.29) were relatively higher than other subscales, while Physical Demand (M=7.83), Temporal Demand (M=12.88), Effort (M=19.44), and Frustration (M=9.09), each ranging between 7--20, were rated more favorably. These results suggest that users did not experience significant physical strain or time pressure while completing tasks using \sysname{}.
The higher Mental Demand and Performance scores were attributed to \sysname{}'s initial version's performance aspects and requirement for users to divide tasks into separate inputs, creating additional cognitive load. P5 expressed this challenge: \textit{"It would be nice to be able to input with such precise divisions, but even this can be difficult for non-developers like me."} P3 also commented: \textit{"This task(Task 2) feels simple enough that it could be done with just ChatGPT instead of using \sysname{}."}

% === Task별 NASA-TLX 하위항목 + Overall 평균 (0~100) ===
%   task  mental_mean  physical_mean  temporal_mean  performance_mean  \
% 0   t1    34.090909       9.090909      18.939394         21.212121   
% 1   t2    22.727273       6.060606      10.606061         49.242424   
% 2   t3    18.181818       8.333333       9.090909         17.424242   

%    effort_mean  frustration_mean  overall_mean  
% 0    22.727273          8.333333     19.065657  
% 1    18.939394          8.333333     19.318182  
% 2    16.666667         10.606061     13.383838  

\begin{table}[t]
\centering
\caption{SUS Scores and Grades for Overall and Individual Tasks}
\label{tab:sus_scores}
\begin{tabular}{lccc}
\toprule
\textbf{Task}   & \textbf{Mean} & \textbf{Std. Dev.} & \textbf{SUS Grade} \\
\midrule
Overall         & 72.65         & 15.07             & Good \\
Task 1          & 74.85         & 16.71             & Good \\
Task 2          & 64.77         & 13.61             & OK   \\
Task 3          & 78.33         & 11.59             & Good \\
\bottomrule
\end{tabular}
\end{table}

SUS scores further validated \sysname{}'s usability.
The overall average SUS score across tasks was 72.65, which corresponds to a ‘Good’ rating based on Bangor et al.’s guidelines~\cite{bangor2008empirical}. Individually, Task 1 received a ‘Good’ rating (74.85), Task 2 an ‘OK’ rating (64.77), and Task 3 a ‘Good’ rating (78.33).
A breakdown of SUS items revealed that Consistency (M=80.05), No learn (M=78.28) and Intuitive (M=77.78) received particularly high ratings, highlighting key strengths of \sysname{} in those aspects. In contrast, Confidence (M=62.12) received comparatively lower scores, indicating areas where the system could be further improved. This nuanced feedback is further supported by qualitative comments from the interviews. For example, participant P4 remarked, “Once you hear the explanation, it is easy enough to use,” which highlights the system’s strong usability. In contrast, participant P8 stated, “Ironically, since it isn’t coding, I still feel uncertain whether it will perform the desired tasks effectively,” pointing to an area that may require further refinement.

\subsection{Improvement of \sysname{}}
The initial study confirmed that \sysname{} delivers satisfactory support for basic task completion. Guided by insights from the initial usability testing—and specifically by the NASA-TLX results indicating higher ratings for the Mental Demand and Performance dimensions—we enhanced the system’s overall usability. To address these issues, we improved system stability and response times and introduced a planning agent that enables lengthy prompt inputs to be split. Based on the SUS findings—particularly concerning the Confidence item—we refined our strategy by not only exposing the chat interface but also by presenting relevant data and actions (see Figure~\ref{fig:aiap1} (d), (e)).

Furthermore, we expanded \sysname{}’s capabilities to support a wider range of tasks beyond the predefined ones, thereby empowering users to create personalized workflows. This enhancement involved incorporating greater flexibility, more diverse input types, and broader support for varied usage scenarios, laying the foundation for the second phase of our study. In addition, we refined the orchestration of the system's multi-agent framework to improve overall efficiency.

\subsection{Study 2: Free-Exploration User Experience Study}
Following the improvements made to \sysname{} based on findings from the initial study, we conducted a second user study to examine how users naturally interact with \sysname{} in open-ended scenarios. This free-exploration study aimed to evaluate not only usability but also the broader user experience by allowing participants to design AI services without predefined instructions or constraints.

\subsubsection{Participants}
We recruited an additional 10 participants who had not taken part in Study 1, ensuring that prior exposure to \sysname{} would not influence their behavior or feedback. As in the previous study, participants included 5 professionals and 5 graduate students, all of whom had basic familiarity with AI. Participants were compensated \$20 per session for their time and insights.

\subsubsection{Procedure and Tasks}
Each session was conducted individually, with participants interacting one-on-one with the researcher. Sessions lasted approximately one hour and were audio- and screen-recorded following the collection of informed consent.

Before beginning the tasks, participants received a brief introduction to \sysname{}, but no specific demonstrations were provided. This was done to ensure that all interactions reflected their natural, first-time use of the tool. Participants were then asked to freely design and implement an AI service of their choice using \sysname{}.

To support this, participants were instructed in advance to come prepared with ideas for an AI service they would like to create. This allowed us to observe how \sysname{} supports users in realizing their unique, personalized workflows.

After completing their task, participants filled out the User Experience Questionnaire (UEQ)~\cite{laugwitz2008construction}. Unlike SUS or NASA-TLX, which focus primarily on usability and workload, the UEQ is a more comprehensive instrument that captures a wide range of user experience factors, including emotional and aesthetic responses. It consists of six bipolar dimensions: Attractiveness, Perspicuity, Efficiency, Dependability, Stimulation, and Novelty. Participants rated each item on a 7-point scale from -3 to +3. These dimensions also guided the structure of follow-up interviews to obtain deeper qualitative insights.

Additionally, we conducted a post-hoc interview to complement the survey. Specifically, we asked questions about four features of the tool and how the multi-agent approach helped them to complete their tasks. 

\begin{figure}[!ht]
      \centering
      \includegraphics[width=0.40\textwidth]{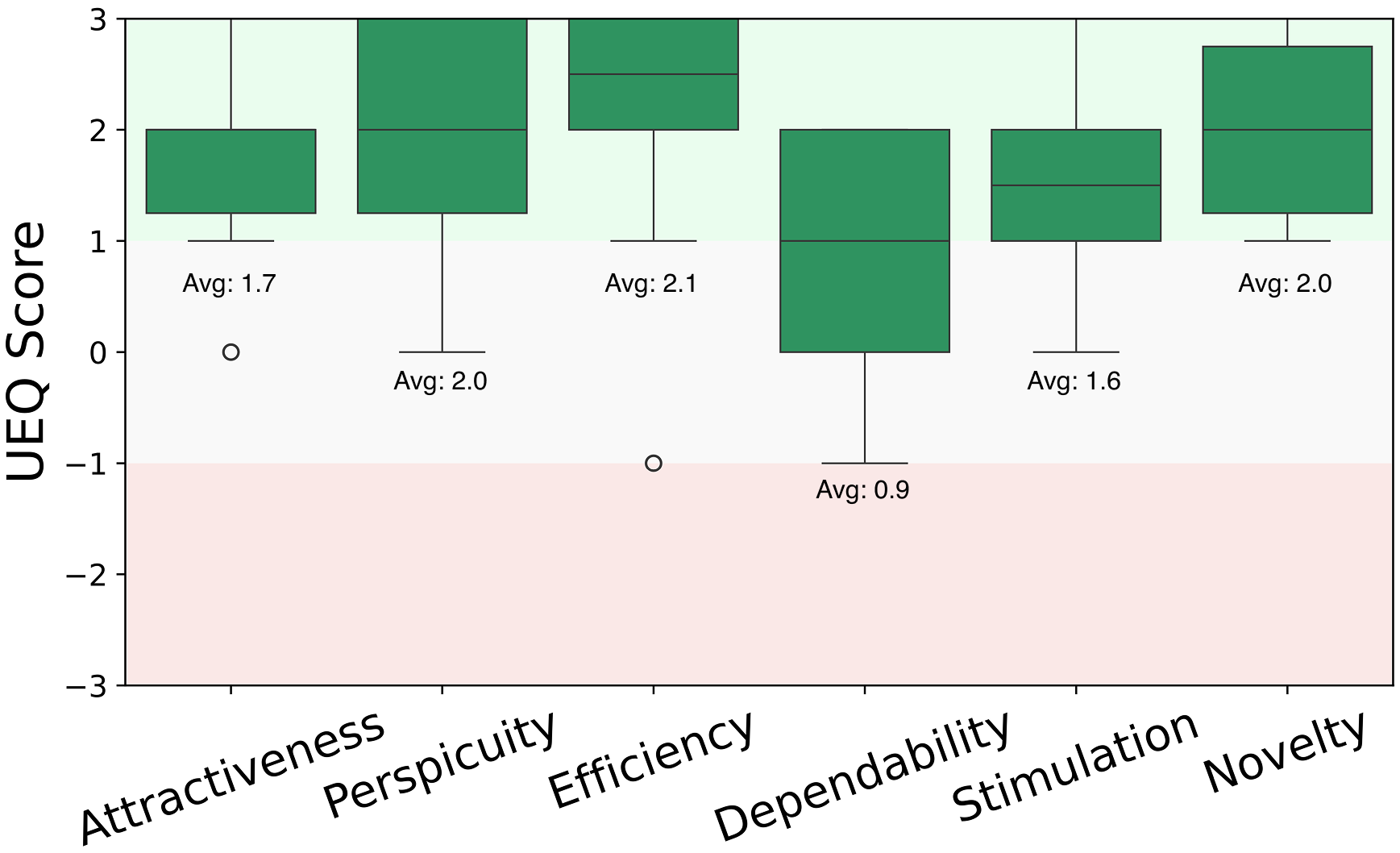}
      \caption{UEQ results of the study. Efficiency scored highest overall, while Dependability scored lowest, reflecting trust-related concerns.}
      \Description{UEQ results of the study. Efficiency scored highest overall, while Dependability scored lowest, reflecting trust-related concerns.}
      \label{fig:ueq}
\end{figure}

\subsubsection{Results}

{\textbf{UEQ:}}
Analysis of UEQ responses revealed that Efficiency received the highest score, with an average of 2.1, indicating strong user perceptions of the system’s ability to support fast and effective task completion. Most participants echoed this sentiment in interviews (e.g., P23, P24, P26, P30), noting that they were able to carry out their envisioned services quickly and without friction. However, some participants suggested areas for improvement. For example, P32 rated Efficiency at -1 and remarked: \textit{"To be more efficient, the system should go beyond simply optimizing design and enable problem-solving through reasoning. For example, instead of analyzing sales data and sending it via email, it should generate workflows suggesting how to increase sales."} This points toward future directions for evolving \sysname{} into a more intelligent, recommendation-driven platform.

Scores for Novelty and Attractiveness were also high, averaging 2.0 and 1.7 respectively, while Stimulation received a mean score of 1.6. Participants consistently described \sysname{} as visually distinct and aesthetically engaging, noting that the block- and flow-based interface felt refreshingly different from conventional LLM services like ChatGPT. They appreciated both the visual design and the modular node interface, highlighting that \sysname{} delivered a more immersive and design-oriented user experience.

In contrast, Dependability received the lowest average score of 0.9, suggesting some concerns regarding the reliability and predictability of system behavior. For example, P25 commented, \textit{"Even with nearly identical inputs, the outputs varied slightly—probably due to the nature of LLMs."} Similarly, P29 noted, \textit{"It doesn’t feel like it provides precise answers, as one would expect for tasks like mathematics or coding. The idea of building logic with LLM feels somewhat awkward."} These insights underscore the importance of improving consistency and reliability in LLM-driven systems, especially when they are used as tools for building functional services.

\textbf{{Qualitative Findings:}}
Beyond the quantitative results, our observational and interview data revealed several additional insights. While participants did not explicitly mention \sysname{}'s multi-agent collaboration—likely because it operated automatically in the background and was not directly perceivable—they offered concrete feedback on various features of the system that shaped their overall experience.

First, participants reported that \sysname{}’s diverse AI functionalities supported them effectively throughout the task execution process. In particular, they expressed satisfaction with the system’s \textit{AI-Generated Suggestions}. These suggestions reformulated user-entered queries into a clearer form based on the system’s interpretation and prompted users to confirm before proceeding. This verification step helped participants intuitively recognize that the system had correctly understood their intent.
Furthermore, this step-by-step interaction style contributed to a positive reception of \sysname{}'s module sequential structure. P17 remarked, \textit{"\sysname{} is easy to understand because it displays processes step by step."} P7 echoed this, stating, \textit{"You can input commands intuitively, step by step, and confirm results at each stage, making it straightforward and easy to follow."} Others (e.g., P23) described the flow as natural, likening it to a conversation with ChatGPT. As P31 noted, \textit{"Prompts work in one direction, and \sysname{} follows the same concept of one-way input and output, making it easier to understand."} These comments underscore the effectiveness of \sysname{}’s unidirectional design in promoting simplicity and predictability—qualities especially important when developing moderately complex but intuitive services.

Second, participants praised the modular structure of \sysname{}. P18 noted, \textit{"It was convenient to rearrange the order of the prompts I entered easily. If I missed something or added something incorrectly, I could just change the sequence or edit that part."} This flexibility allowed users to iteratively refine their workflows without starting from scratch. P27 added, \textit{"The modular structure lets you set a basic framework and then freely add or reorder prompts, making it great for transparently seeing how the service operates and reviewing it step by step as needed."}
% P2 even compared the system to ComfyUI~\cite{comfyui}, stating, \textit{"It reminded me of ComfyUI. It’s easy to understand because each step is separated and executed individually."} 
This modularity was widely appreciated for its ability to support transparency, reusability, and iterative thinking.

Participants also emphasized the benefits of \sysname{}’s \textit{Automatic Identification of Data, Action, and Context}. By breaking down instructions into key components—Data (typically nouns), Actions (verbs), and Context (additional descriptive elements)—participants reported that they were able to gain a more granular and intuitive understanding of the elements required to complete a task. This structured parsing helped users think through their goals in a more systematic way, often surfacing elements they may have otherwise overlooked.
In particular, this feature was seen as effective in mitigating common issues of data omission. Participants noted that the system’s ability to detect missing or incomplete inputs contributed to more complete and accurate outcomes. For example, P20 remarked, \textit{"ChatGPT often provides responses without core information, which I suspect might be hallucination."} Another participant noted, \textit{"Compared to just entering prompts, it looks more complex with more information, but it’s definitely clearer."} Participants appreciated how \sysname{} not only automatically linked prompts to relevant APIs, but also proactively surfaced essential missing components. As P5 explained, \textit{"It was nice to focus only on what I wanted to do without worrying about how to implement it or which API to use."} While this level of automation was particularly helpful for novice users, several more experienced participants expressed a desire for greater control over configuration details and technical parameters. This reflects a recurring design tension between simplicity and flexibility—one that remains central to \sysname{}’s continued evolution.

Regarding specific capabilities, participants also expressed satisfaction with \sysname{}’s \textit{Intelligent Action Linking}. This feature was seen as significantly extending the system’s functional scope. Participants anticipated that it could enable a broader range of real-world, practical applications—especially in professional or task-oriented settings.

In addition to feature-specific feedback, an interesting broader finding was the diversity in prompting styles observed during system use. Some participants (e.g., P23) used short, sequential prompts, comparing the experience to service composition tools like Make: \textit{"This feels similar to a tool I’ve used before, called Make. That tool also requires breaking tasks into separate steps to create them sequentially."} Others (e.g., P26) submitted longer, more descriptive prompts, treating \sysname{} like a conventional LLM: \textit{"I entered a long instruction because it asked for input, but I felt a bit uneasy. It felt like I was providing an open-ended input rather than a precise specification for designing a service."}
Still others (e.g., P31) relied on structured, long-form prompts shaped by their prior experiences with LLMs: \textit{"When I saw the input field, I thought I could use it similarly to an LLM service. I entered the prompt as I usually do when using an LLM service. This generally produces clearer results."}
Interestingly, several participants adapted and refined their prompting styles over time, suggesting a process of learning and skill development through repeated interactions with the system. This evolution points to a potential community-driven growth effect, whereby users build personalized strategies for interacting with \sysname{} and, over time, share those strategies with others—ultimately fostering a collaborative knowledge ecosystem around the tool.

\section{Discussion}
The advancement of AI technology has enabled non-experts to develop sophisticated AI services without requiring complex programming knowledge. The system introduced in this study, \sysname{}, builds on this trend by offering an interface that combines natural language input, visual programming, and multi-agent collaboration. Through this design, \sysname{} presents key implications for the HCI community—especially in terms of interaction design, cognitive scaffolding, and agent-user abstraction strategies. This section reflects on the broader significance of our findings, synthesizes design principles drawn from user behavior, and discusses limitations that future work may address.

\subsection{Harmonizing User Intent and System Understanding}
One of the most persistent challenges non-experts face when interacting with AI systems is articulating their intent clearly while grasping the system’s true capabilities~\cite{subramonyam2024bridging}. Users often default to intuition, heuristics, or trial-and-error strategies~\cite{johnny}, rather than employing systematic planning. This results in what Subramonyam et al.~\cite{subramonyam2024bridging} term the “intentionality gap,” where users struggle to align their high-level goals with system behaviors and outputs.

\sysname{} addresses this gap through a combination of visual programming and multi-agent collaboration. In particular, the AI-Generated Suggestion mechanism functions not only as an autocomplete tool, but also as a cognitive mediator—it reinterprets loosely structured prompts and transforms them into clear, sequenced operations. This interaction style builds on and extends the concept of “co-envisioning”~\cite{subramonyam2024bridging}, in which systems actively participate in goal clarification rather than passively responding to input. By surfacing intermediate representations and prompting user confirmation, \sysname{} helps users develop a more accurate understanding of their own goals and the system's capabilities.

Moreover, this clarification process contributes to user confidence, particularly for those with limited AI or technical experience. Participants frequently reported that the intermediate suggestion phase acted as a “thinking partner,” helping them structure ideas more clearly and validate system comprehension before committing to actions.

\subsection{Structural Approaches to Reducing Cognitive Load}
The data-action-context structure embedded in \sysname{} mirrors natural language composition and supports intuitive reasoning. Unlike traditional visual programming tools that require familiarity with abstract programming concepts, this structure uses linguistically aligned categories—nouns as data, verbs as actions, and phrases as context—to reduce mental friction. By building on these linguistic affordances, \sysname{} allows users to operate within familiar cognitive frameworks, bridging the gap between everyday language and computational logic.

Our user studies reinforce the effectiveness of this strategy. Participants reported that the structured decomposition of prompts helped them better understand system expectations, clarify missing components, and anticipate execution outcomes. In particular, the visual modularization of workflows supported explainability and step-wise verification—two factors closely tied to user trust in intelligent systems~\cite{liao2021human}. For example, participants mentioned that they felt more “in control” when they could visually inspect how each step contributed to the overall workflow, even if they had no technical background.

This structure also served as a useful pedagogical scaffold for participants who were new to service composition. Several users adapted their prompting behavior over time, shifting from open-ended descriptions to more structured inputs aligned with the data-action-context model. This finding suggests that the system not only reduces cognitive load in the moment, but also helps users build transferable mental models for future interactions.

\subsection{Integration Effects of Natural Language and Visual Programming}
Our findings strongly support the use of hybrid interfaces that integrate natural language input with visual workflow representation. This integration leverages the strengths of both modalities: the flexibility and accessibility of natural language, and the structure and transparency of visual workflows. Instead of choosing between ease of expression and formal rigor, \sysname{} enables users to benefit from both simultaneously.

Such interfaces offer more than just usability improvements—they function as cognitive amplifiers. In our study, participants reported that the ability to "speak freely" through language while "seeing clearly" through visual structure allowed them to engage more deeply with their goals and the AI’s interpretation of them. The back-and-forth interplay between verbal intention and visual representation reinforced learning and reduced the need for mental translation.

Importantly, this hybrid interface also contributes to error recovery and iterative refinement. Participants often corrected misunderstandings or rephrased inputs after seeing how the system had structured their requests. The visual layer thus serves as a continuous feedback channel, helping users build mental models not only of the task but of the system’s logic itself.

We conceptualize this integration as part of a broader shift in HCI—from interface-as-instruction to interface-as-dialogue—where AI systems move beyond passive input receivers to become collaborative partners that assist users in forming, refining, and executing their goals. Rather than requiring users to adapt to rigid system protocols, this paradigm emphasizes mutual adaptation, iterative clarification, and shared intentionality between human and machine.

\subsection{A Multi-Agent Service without Perceivable Agents}

\sysname{} employs a multi-agent architecture in which five specialized agents engage in role-specific, multi-turn collaboration. However, this internal orchestration is deliberately abstracted away from users. From the user’s perspective, the system presents a single, unified interface without requiring explicit awareness or control over individual agents. This abstraction reduces cognitive overhead and allows users to focus solely on task-related goals rather than system mechanics.

This design contrasts with systems that expose agent-specific roles or require users to manually configure agent interactions. While such transparency may appeal to expert users, it risks overwhelming non-experts with unnecessary complexity, particularly when they must reason about delegation, responsibility, or inter-agent dependencies.

To address this, we propose the concept of a "one-agent user experience (one-agent UX)"—a design paradigm in which multiple backend agents are coordinated to present a consistent, coherent, and unified interaction layer. Although agent-based collaboration occurs internally, the system behaves in a manner that is functionally holistic and perceptually singular. This approach maintains transparency at the task level while hiding architectural complexity, aligning with prior research on interaction abstraction and cognitive alignment~\cite{weld2019challenge}.

A useful metaphor is found in the film Inside Out, where internal emotional agents collectively influence behavior, yet the character Riley presents as a single, cohesive persona. Similarly, users of \sysname{} benefit from features such as suggestion, linking, validation, and refinement—without needing to comprehend or coordinate the underlying agentic processes.

Observations from our user study support this principle: participants perceived the system as unified and consistent, despite the complex orchestration behind the scenes. This suggests that multi-agent intelligence can be most effective when it is behaviorally coherent, contextually adaptive, and structurally unobtrusive. We recommend the one-agent UX as a guiding principle for designing future multi-agent systems that prioritize user accessibility, trust, and clarity.

\section{Limitations and Future Work}

We acknowledge several limitations in our study. First, our user studies were conducted in limited experimental settings over a short period, constraining our ability to fully validate the system's effectiveness across diverse real-world situations and user contexts. This restricted our observation of unexpected issues and long-term usage patterns that might emerge in actual work environments.
Second, despite the \sysname{}'s structured approach, some users initially struggled to grasp the concept of Data-Action-Context segmentation. While AI Suggestions helped mitigate this challenge, further refinements in onboarding design may be needed to lower the initial learning curve.
Third, our research primarily focused on text-based LLMs, without exploring the potential applicability to other modalities such as visual language models for image or video generation.
Future research directions include evolving the AI workflow builder beyond user-intent design into a comprehensive transparency management tool for generative AI agents. Additionally, exploring the platform's potential applicability to multimodal scenarios, such as image and video workflows, represents an important direction for future development. Furthermore, we plan to enhance the system's applicability and scalability across more diverse environments by leveraging MCP (Model Context Protocol)\footnote{\url{https://docs.anthropic.com/en/docs/agents-and-tools/mcp}} and emerging technical protocols.
% 이 논문은 사용성을 위주로 다루었고 추후 성능에 대한 검증을 다룰 예정이다.?

\section{Conclusion}

This study introduced \sysname{}, a no-code AI platform specifically designed to bridge the existing gaps between user intent and AI systems. By seamlessly integrating a natural language interface with visual workflows and internally orchestrating a transparent multi-agent collaboration, \sysname{} enables non-experts to intuitively design personalized AI services. In particular, it effectively addresses well-documented challenges in HCI research, such as intentionality gaps, cognitive gaps, and technical gaps, through a structured pipeline and clear interface design (e.g., AI-generated suggestions, modular workflows, and Data-Action-Context decomposition).

Two phases of user studies validated the strengths of \sysname{}, highlighting its efficiency, flexibility, and intuitiveness, and users reported a generally positive user experience. However, some participants pointed out that for extremely simple tasks, such as basic translations, traditional chat-based services might offer greater convenience. Others noted ambiguity in using natural language to define workflows, suggesting potential uncertainty regarding task execution. This feedback indicates a need for selectively incorporating certain characteristics from traditional development tools to enhance clarity and user confidence, especially in moderately complex scenarios.

Despite these limitations, \sysname{} successfully combines the strengths of natural language-based workflow builders and visual programming methods to effectively resolve critical interaction issues. This research thus contributes meaningful directions and practical design principles for future user-centered AI tool design in the HCI field.

\bibliographystyle{ACM-Reference-Format}
\bibliography{ref}

\end{document}